\documentclass[12pt,preprint]{aastex}
\usepackage{graphicx} 

\slugcomment{Not to appear in Nonlearned J., 45.}

\shorttitle{magnetic helicity}
\shortauthors{Zhang et al.}

\begin{document}


\title{Magnetic helicity transported by flux emergence and \\shuffling motions in Solar Active Region NOAA 10930}


\author{Y. Zhang\altaffilmark{1,2}, R. Kitai\altaffilmark{1}, and K. Takizawa\altaffilmark{1}}
\affil{\altaffilmark{1}Kwasan and Hida Observatories, Kyoto University, Yamashina-ku, Kyoto 607-8471}

\affil{\altaffilmark{2}Key Laboratory of Solar Activity, National Astronomical Observatories, Chinese Academy of Sciences, Chaoyang distribution, Beijing 100012}
\email{zhangyin@kwasan.kyoto-u.ac.jp\\ zhangyin@bao.ac.cn}

\begin{abstract}

We present a new methodology which can determine magnetic helicity transport by the passage of helical magnetic field lines from sub-photosphere and the shuffling motions of foot-points of preexisting coronal field lines separately. It is well known that only the velocity component which is perpendicular to the magnetic field ($\mbox{\boldmath$\upsilon$}_{\perp B}$) has contribution to the helicity accumulation. Here, we demonstrate that $\mbox{\boldmath$\upsilon$}_{\perp B}$ can be deduced from horizontal motion and vector magnetograms, under a simple relation of $\mbox{\boldmath$\upsilon$}_{t}=\mbox{\boldmath$\mu$}_{t}+\frac{\upsilon_{n}}{B_{n}}\textbf{\textit{B}}_{t}$ as suggested by D$\acute{e}$moulin \& Berger (2003). Then after dividing $\mbox{\boldmath$\upsilon$}_{\perp B}$ into two components, as one is tangential and the other is normal to the solar surface, we can determine both terms of helicity transport. Active region (AR) NOAA 10930 is analyzed as an example during its solar disk center passage by using data obtained by the Spectro-Polarimeter and the Narrowband Filter Imager of Solar Optical Telescope on board Hinode. We find that in our calculation, the helicity injection by flux emergence and shuffling motions have the same sign. During the period we studied, the main contribution of helicity accumulation comes from the flux emergence effect, while the dynamic transient evolution comes from the shuffling motions effect. Our observational results further indicate that for this AR, the apparent rotational motion in the following sunspot is the real shuffling motions on solar surface.

\end{abstract}


\section{Introduction}

It is widely believed that eruptions in solar atmosphere, such as flare, filament eruption and coronal mass ejection (CME), are the intermittent liberation of non-potential magnetic energy stored in coronal magnetic field. Magnetic helicity is a quantitative measure of twists, kinks, and inter-linkages of magnetic field lines (Berger \& Field 1984) and is a useful and important parameter to indicate topology and non-potentiality of a magnetic field system. Moreover, it is conserved in a closed volume, as well as in an open volume in the absence of boundary flows (Berger \& Field 1984). Since the solar corona is an open volume with the photosphere as a boundary with normal flux, the magnetic helicity can be transported from the sub-photosphere into the corona though the boundary of photosphere by emergence of helical flux and shuffling motions (Berger \& Field 1984). According to Berger (1999), the transport rate of relative helicity, $\dot{H}$, due to both processes are given by

\begin{equation}
\dot{H_{n}}=\oint2(\textbf{\textit{B}}_{t} \cdot
    \textbf{\textit{A}}_{{p}}){\upsilon_{n}}dS
\end{equation}
\begin{equation}
  \dot{H_{t}}=-\oint2(\mbox{\boldmath$\upsilon$}_{t}\cdot \textbf{\textit{A}}_{{p}})B_{{n}}dS
\end{equation}

where subscripts \textquotedblleft n\textquotedblright and \textquotedblleft t\textquotedblright represent the normal and tangential component, respectively. Chae (2001) has observationally determined $\dot{H_{t}}$ by regarding the horizontal motions $\mbox{\boldmath$\mu$}_{t}$ which deduced from a time series of line-of-sight magnetograms with the local correlation tracking (LCT) method (November \& Simon 1988) as the shuffling motions $\mbox{\boldmath$\upsilon$}_{t}$, and found that shuffling motions on photosphere can provide enough magnetic helicity for the CMEs, which was confirmed by a series of subsequent studies (Nindos \& Zhang 2002; Moon et al. 2002a,b; Nindos et al. 2003). It also has been proved that in some active regions (ARs), magnetic helicity transported by rotational motion is comparable with that deduced from LCT method (Zhang et al., 2008; Min et al., 2009). However, D$\acute{e}$moulin \& Berger (2003) have demonstrated that the horizontal motions $\mbox{\boldmath$\mu$}_{t}$ include both components $\mbox{\boldmath$\upsilon$}_{t}$ and $\mbox{\boldmath$\upsilon$}_{n}$ with the relation of $\mbox{\boldmath$\mu$}_{t}=\mbox{\boldmath$\upsilon$}_{t}-\frac{\upsilon_{n}}{B_{n}}\textbf{\textit{B}}_{t}$. Therefore, helicity injection calculated this way not only include the contribution of shuffling motions but also the emerging fluxes. 

Emergence of twisted fluxes and shuffling motions are the popular triggers for solar eruptions and the common mechanism for magnetic helicity accumulation both in theoretical and observational works (Leka et al. 1996; Grigoryev \& Ermakova 2002; Fan \& Gibson 2004). Emergence naturally carries magnetic flux as well as helicity though the photosphere, if the emerging fluxes have magnetic helicity. While shuffling motions generate magnetic shear and then supply helicity and free energy into coronal field. But still now, we know a little about to what extent and how both processes contribute to the helicity accumulation. 

It is worth noting that the presence or magnitude of flow along the field lines has no bearing on the temporal evolution of the magnetic field (as described by the ideal MHD induction equation), and also has no contribution to the helicity accumulation. So we should consider only the motion which is perpendicular to the magnetic field in the estimation of helicity accumulation. D$\acute{e}$moulin \& Berger (2003) have pointed out that there are two ways to get the normal velocity, by which we can calculate the magnetic helicity transport by flux emergence and shuffling motions separately. Firstly, since the magnetic field and velocity field are not mutually independent, we can deduce the normal velocity from known conditions under some assumptions. Secondly, if the AR locates around the center of solar disk, we can use Doppler velocity as the normal velocity. Advantages and disadvantages of both methods were discussed by many authors (D$\acute{e}$moulin \& Berger 2003; Pariat et al. 2005). So far, only the first method has been tried. Kusano et al. (2002) have deduced $\upsilon_{n}$ and $\mbox{\boldmath$\upsilon$}_{t}$ which only include component perpendicular to the magnetic field by using observations of $\textbf{\textit{B}}$ and $\mbox{\boldmath$\mu$}_{t}$ on photosphere together with the induction equation. They found that both processes have equal contribution to the helicity injection and have supplied magnetic helicity of opposite signs in AR NOAA 8210. Welsch et al. (2004) have introduced two methods to get the normal velocity. One is to deduce the $\mbox{\boldmath$\upsilon$}_{t}$ and $\mbox{\boldmath$\upsilon$}_{n}$ under an assumption of $\mbox{\boldmath$\upsilon$}\cdot\textbf{\textit{B}}=0$ and by a simple relation $\mbox{\boldmath$\upsilon$}_{t}=\mbox{\boldmath$\mu$}_{t}+\frac{\upsilon_{n}}{B_{n}}\textbf{\textit{B}}_{t}$ proposed by D$\acute{e}$moulin \& Berger (2003), and the other is to solve the same equations as Kusano et al. (2002) by different technique. Using the high spatial resolution vector magnetogram obtained by Helioseismic and Magnetic Imager (HMI) on broad the Solar Dynamics Observatory (SDO), Liu et al. calculated the velocity vector by the differential affine velocity estimator (DAVE) method, which models image motion with either the continuity equation or the convection equation (depending on a switch). And then they found that the helicity flux from shuffling term is dominant, while that from emergence term is small (private communication).

Based on a simple relation  $\mbox{\boldmath$\upsilon$}_{t}=\mbox{\boldmath$\mu$}_{t}+\frac{\upsilon_{n}}{B_{n}}\textbf{\textit{B}}_{t}$ (D$\acute{e}$moulin \& Berger 2003), we find that the component of velocity which is perpendicular to the magnetic field can be deduced from the vector magnetograms and the horizontal motions (see detail in section 3). After projecting this velocity into tangential and normal component of photosphere, we can deduce the magnetic helicity transport by shuffling motions and flux emergence separately. We apply our new method to an isolated AR NOAA 10930. In order to decrease the ambiguities, we only study the day that the AR passed through the solar meridian. The paper was organized as follows. Section 2 is the description of the observations. The method of how to get the plasma velocity which is perpendicular to the magnetic field and data reduction are presented in Section 3. Section 4 are the observational results. Summary and discussion are presented in Section 5.

\section{Observations}

Figure 1 shows the morphological evolution of AR NOAA 10930 from December 08 to 13. It has a mature and stable leading sunspot, and a small and rapidly changed following sunspot. The temporal evolution of the sunspots area (thick (thin) line for leading (following) sunspot) and tilt angle are also shown in Figure 1. In order to distinguish opposite polarities in $\delta$ sunspot, only pixels whose intensity is weaker than 45\% of quiet sun are included for area estimation. The tilt angle is measured as angle between the connection of two polarities and the south direction. From Figure 1, we see that the leading sunspot was stable during its solar disk passage. The morphology of the following sunspot changed very rapidly according to emergence, cancellation and mergence, whereas its area changed a little until December 9. After that, there was a significant emergence. It slowed down very quickly at the beginning of December 10. Around 12:00 UT on December 10, the flux emergence reoccurred. Associated with the emergence, the following sunspot rotated counterclockwise around its center and moved eastward rapidly, resulting in an increase in magnetic complexity. These growing process continued till a X3.4 flare occurred on December 13.

AR NOAA 10930 passed through the solar meridian on 2006 December 11, with the location changing from W4S6 to E7S6. Meanwhile, the following sunspot area increased from $4.9\times10^{7}$ km$^{2}$ to $8.7\times10^{7}$ km$^{2}$, as a percentage of 35\% of the total area increase during its solar disk passage. The variations of tilt angle which due to the eastward motion of the following sunspot changed from 10$^{\circ}$ to 32$^{\circ}$, as a percentage of 50\% of the total increase of the tilt angle. The rotational speed of following sunspot increased up to $8^{\circ} h^{-1}$ on December 11. It maintained till a X3.4 flare occurred on December 13. And then the rotational speed slowed down to $3^{\circ} h^{-1}$ (Min et al. 2009). All the observational evidences show that December 11 was an important day for the AR evolution. 

  \begin{figure}    
   \centerline{\includegraphics[width=0.6\textwidth,clip=]{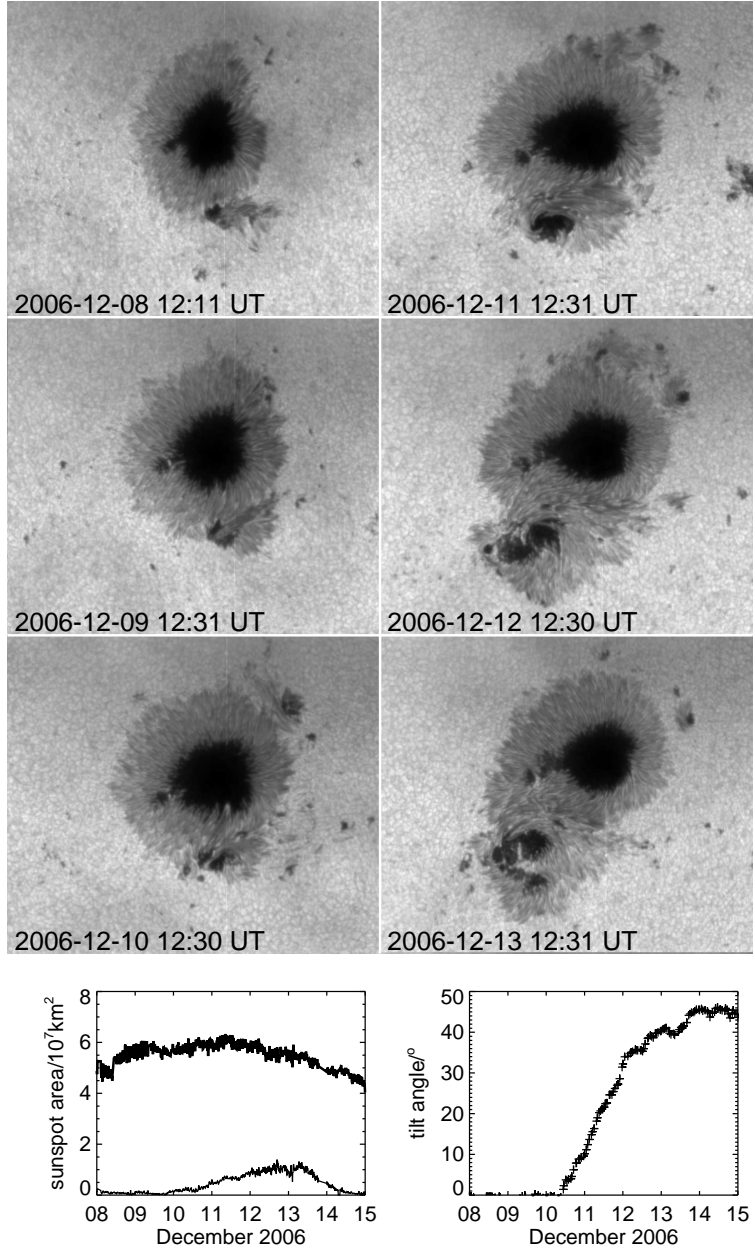}              }
              \caption{Images show AR NOAA 10930 evolution during its solar disk passage from December 8 to 14. Plots in bottom panels show temporal evolution of sunspots area and tilt angle. }
   \end{figure}

\section{Method and Data Reduction}

\subsection{Deduction of Plasma Velocity Perpendicular to the Magnetic Field}

At first, we give a simple description of the relation among different velocities which are used in the present work. From observations, we can get the horizontal motions ($\mbox{\boldmath$\mu$}_{t}$) of the magnetic features by LCT method, which we indicate by purple arrow in Figure 2a. Shuffling motions of plasma on solar surface ($\mbox{\boldmath$\upsilon$}_{t}$) can be deduced from $\mbox{\boldmath$\mu$}_{t}$ by $\mbox{\boldmath$\upsilon$}_{t}=\mbox{\boldmath$\mu$}_{t}+\frac{\upsilon_{n}}{B_{n}}\textbf{\textit{B}}_{t}$. Normal velocity of plasma ($\mbox{\boldmath$\upsilon$}_{n}$) can be obtained by several ways as we introduced in previous section. Both $\mbox{\boldmath$\upsilon$}_{t}$ and $\mbox{\boldmath$\upsilon$}_{n}$ 
are outlined by red arrows in Figure 2a. The combined vector of them is the real motion of plasma ($\mbox{\boldmath$\upsilon$}$), which is represented by black arrow in Figure 2a. With reference to the direction of magnetic field, as shown by thick black arrow in Figure 2a, $\mbox{\boldmath$\upsilon$}$ can be divided into two components, one is parallel to the magnetic field ($\mbox{\boldmath$\upsilon$}_{\parallel}$) and the other is perpendicular to the magnetic field ($\mbox{\boldmath$\upsilon$}_{\perp}$), as we shown in Figure 2a by blue arrows. Since $\mbox{\boldmath$\upsilon$}_{\parallel}$ has no contribution to the magnetic helicity accumulation, we neglect it in the present work. Then, we get the normal velocity and shuffling motions of the plasma by dividing $\mbox{\boldmath$\upsilon$}_{\perp}$ into two components, one is tangential to the solar surface ($\mbox{\boldmath$\upsilon$}_{\perp t}$) and the other is normal to the solar surface ($\mbox{\boldmath$\upsilon$}_{\perp n}$), as we show by green arrows in Figure 2a. According to the vector relations depicted in Figure 2a, we get
 \begin{equation}
\mbox{\boldmath$\upsilon$}_{\perp}=\mbox{\boldmath$\mu$}_{t}-(\mbox{\boldmath$\mu$}_{t}\cdot\textbf{\textit{b}})\textbf{\textit{b}}
\end{equation}
where \textbf{\textit{b}} is the direction vector of magnetic field. That means $\mbox{\boldmath$\upsilon$}_{\perp}$ can be deduced from $\mbox{\boldmath$\mu$}_{t}$ and $\textbf{\textit{B}}$.

\subsection{Data Reduction}

The Spectro-Polarimeter (SP) of Solar Optical Telescope (SOT) on board Hinode measures spectral profiles of full Stokes parameters of two Fe I lines at 630.15 and 630.25 nm and nearby continuum with 21.5 m$\mathring{A}$ spectral resolution (Kosugi et al. 2007; Suematsu et al. 2007; Ichimoto et al. 2007). On 2006 December 11, SP observed AR NOAA 10930 under fast map mode with the spatial resolution of 0.32$''$ and obtained 6 sets of Stokes I, Q, U and V. Vector magnetograms are derived from the inversion of the full Stokes profiles based on the assumption of Milne-Eddington (ME) atmosphere model. We use an automated ambiguity-resolution code based on the minimum energy algorithm (Leka et al. 2009) to resolve the 180$^\circ$ ambiguity of magnetograms. 
The noise level is estimated by selecting one quiet region within the SP map to calculate 1$\sigma$ standard deviation of the average value of field strengths. In order to avoid ambiguity, we only analyze those pixels with $\textbf{\textit{B}}_{n}$ and $\textbf{\textit{B}}_{t}$ greater than the above deviation. Figure 2b shows a vector magnetogram on 11:10 UT. It shows that the arrows in leading sunspot are almost radial except the southwest part, as we outlined by white thick curve and labeled as A. In this region, the magnetic field is obviously left handed. Meanwhile, the following sunspot is also left handed, which corresponds to negative helicity for AR located in south hemisphere. There is so called \textquotedblleft magnetic channel\textquotedblright (Wang et al. 2008) structure along the neutral line. Magnetic field here is highly sheared, as indicated by arrows parallel to the neutral line.

The horizontal motions $\mbox{\boldmath$\mu$}_{t}$ were calculated by 2 minutes longitudinal magnetograms taken by the Narrowband Filter Imager (NFI) of SOT (Tsuneta et al. 2008) by LCT method, and is shown in Figure 2c. Our result shows the similar organized motions, such as inward motion in inner-penumbra and outward motion in outer-penumbra of the leading sunspot, and counterclockwise rotational motion and the eastward motion of the following sunspot, which are consistent with Min et al. (2009). Furthermore, horizontal motions in region A shows clockwise rotation, which correspond to positive helicity injection.

Figure 2d shows $\mbox{\boldmath$\upsilon$}_{\perp t}$ which deduced from $\mbox{\boldmath$\mu$}_{t}$ and \textit{\textbf{B}}. In this map, for the leading sunspot, the inward motion in inner-penumbra remains, while the outward motion in outer-penumbra disappears. Meanwhile, in the following sunspot, the counterclockwise rotational motion remains, while the eastward motion disappears. It means that horizontal motion which deduced by LCT method do include two components. One is the shuffling motion, which is the real horizontal motion of the plasma on the photosphere in more or less vertical magnetic field. It will drag the magnetic field associated with the plasma, and remains in the $\mbox{\boldmath$\upsilon$}_{\perp t}$ map. The other results from the flux emergence or submergence in nearly horizontal magnetic configuration and then footpoints of tubes move along the magnetic field. This motion will disappear from the $\mbox{\boldmath$\upsilon$}_{\perp t}$ map. As in our example, the outward motion in out-penumbra in leading sunspot and the eastward motion along the neutral line disappear from the $\mbox{\boldmath$\upsilon$}_{\perp t}$ map, which are motions resulting from the emergence since the sunspot is going through its growing phase. Meanwhile, the inward motion in inner-penumbra of leading sunspot and the counterclockwise rotation in following sunspot, which remain in $\mbox{\boldmath$\upsilon$}_{\perp t}$ map, are the shuffling motions of the plasma. Furthermore, in this map, the clockwise rotation in region A is more outstanding than that in $\mbox{\boldmath$\mu$}_{t}$ map. 

The normal velocity, $\mbox{\boldmath$\upsilon$}_{\perp n}$, is shown in Figure 2e, where white tone represents upward motion and black one represents downward motion. It shows that the sunspot is dominated by upward motion. But at the boundary of penumbra and photosphere, the upward and downward motions are mixed, which presents a complex distribution of normal velocity. As a comparison, Doppler map, which is calculated from spectral line core fits to Fe I 630.15 nm line profile at each spatial position, is shown in Figure 2f. The pixel resolution of $\mbox{\boldmath$\upsilon$}_{\perp n}$ map is much lower than that of Doppler map, because it was deduced from $\mbox{\boldmath$\mu$}_{t}$ map. So more small features can be identified in Doppler map. However, roughly speaking both maps are consistent. Especially, there are some downward motions along the neutral line. Compared Figure 2e and 2f to 2b, we find that these down flow area are spatially corresponds to the magnetic channel. 

 \begin{figure}    
   \centerline{\includegraphics[width=0.8\textwidth,clip=]{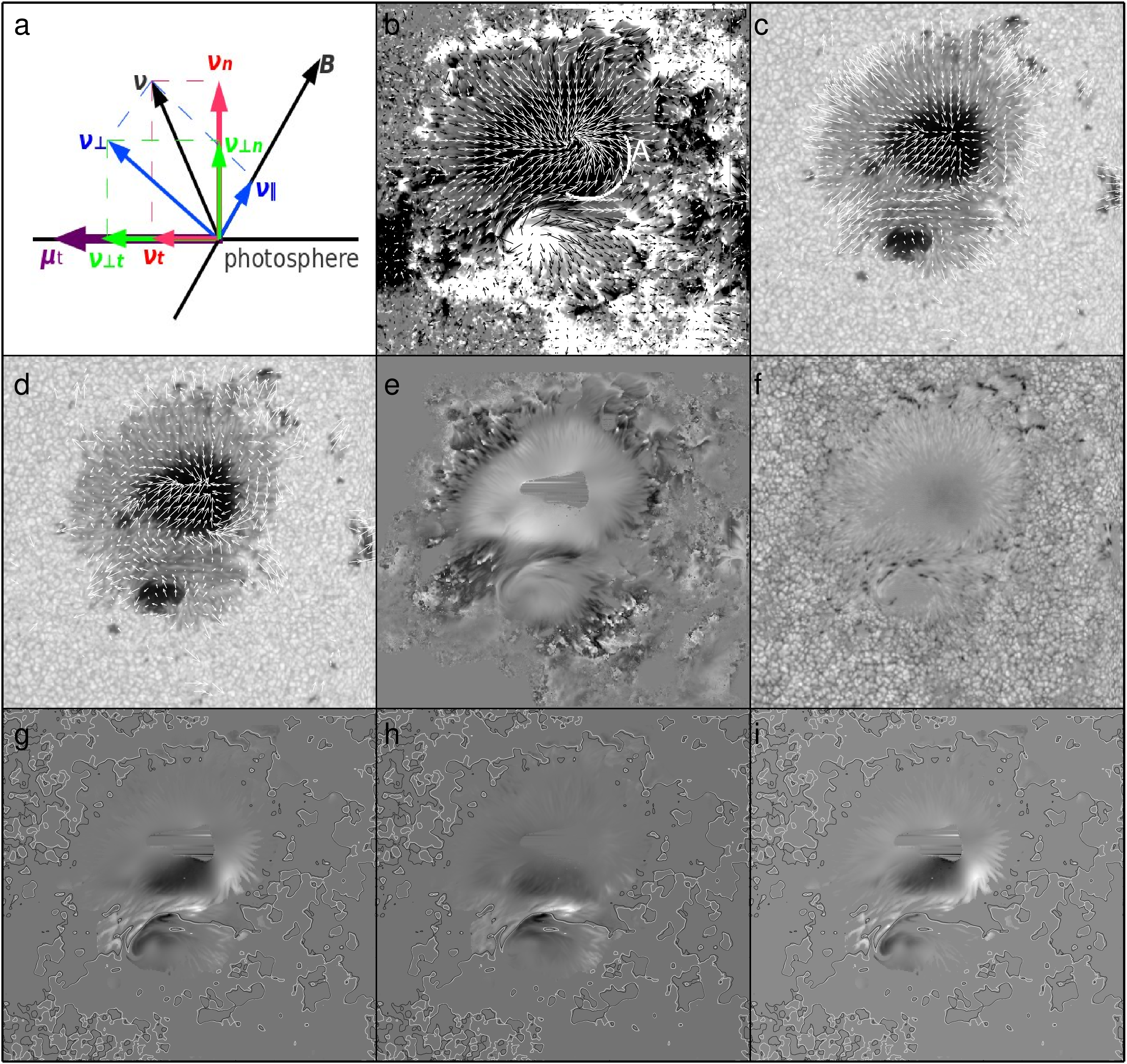}              }
              \caption{(a)A sketch to show the relation of different velocities which used in the present work. Purple arrow represents the horizontal motion $\mbox{\boldmath$\mu$}_{t}$. Red arrows represent the real motion of the plasma on solar surface $\mbox{\boldmath$\upsilon$}_{t}$ and the normal component of plasma motion $\mbox{\boldmath$\upsilon$}_{n}$. The combined vector of both is the real motion of the plasma ($\mbox{\boldmath$\upsilon$}$) as indicated by thin black arrow. With reference to the direction of magnetic field, as shown by thick black arrows, $\mbox{\boldmath$\upsilon$}$ can be divided into two components. One is parallel to the magnetic field ($\mbox{\boldmath$\upsilon$}_{\parallel}$) and the other is perpendicular to the magnetic field ($\mbox{\boldmath$\upsilon$}_{\perp}$), as we shown by blue arrows; (b) Vector magnetogram; (c) $\mbox{\boldmath$\mu$}_{\perp t}$; (d) $\mbox{\boldmath$\upsilon$}_{\perp t}$; (e) $\mbox{\boldmath$\upsilon$}_{\perp n}$; (f) Doppler velocity; (g) Map of helicity flow, $2(\textbf{\textit{B}}_{t} \cdot              \textbf{\textit{A}}_{{p}}){\upsilon_{\perp n}}-2(\mbox{\boldmath$\upsilon$}_{\perp t}\cdot \textbf{\textit{A}}_{{p}})B_{{n}}$ (h) Map of helicity flow due to the vertical motion, $2(\textbf{\textit{B}}_{t} \cdot              \textbf{\textit{A}}_{{p}}){\upsilon_{\perp n}}$; (i) Map of helicity flow due to the shuffling motion $-2(\mbox{\boldmath$\upsilon$}_{\perp t}\cdot \textbf{\textit{A}}_{{p}})B_{{n}}$
                      }
   \end{figure}

\begin{figure}  
   \centerline{\includegraphics[width=0.4\textwidth,clip=]{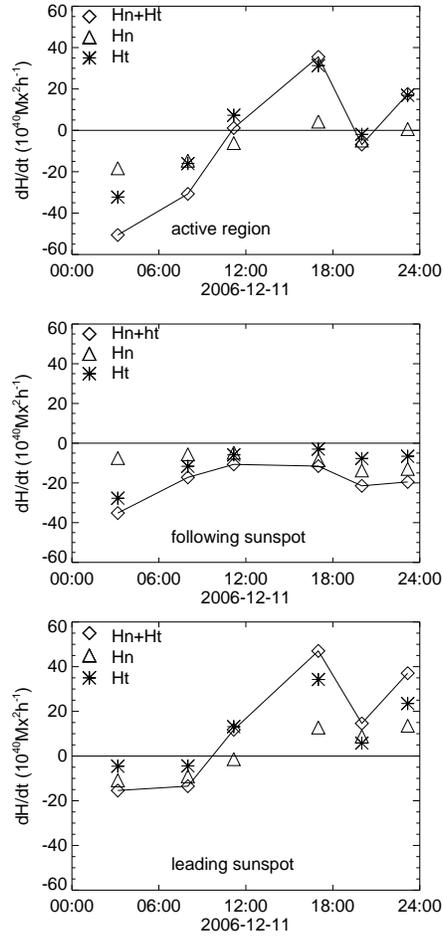}              }
              \caption{Injection rates of helicity due to flux emergence (triangle signs), shuffling motions (star sings), and sum of both (diamond signs) for whole AR, following sunspot and leading sunspot, respectively.
                      }
   \end{figure}

\section{Results}

Form Zhang et al. (2008) \& Park et al. (2010), we know the magnetic helicity injection in AR NOAA 10930 is predominantly negative throughout the period of its disk passage. Park et al. (2010) found that there are three time periods over which magnetic helicity injection is mainly positive for more than nine hours. The day we studied in the present work is within the second period of the positive helicity injection.

Figure 2g shows the distribution of $-2(\mbox{\boldmath$\upsilon$}_{\perp t}\cdot \textbf{\textit{A}}_{{p}})B_{{n}}dS+2(\textbf{\textit{B}}_{t} \cdot\textbf{\textit{A}}_{{p}}){\upsilon_{\perp n}}dS$ on 11:10 UT, which is a sum of local contribution by flux emergence and shuffling motions. It shows that the helicity injection in following sunspot is predominantly negative, which is consistent with the left handed twisted magnetic field and counterclockwise rotation of the sunspot. While in leading sunspot, both negative and positive helicity injection occurred. It is clearly seen that the positive helicity injection occurred at two separate regions. One is along the neutral line. The other is the region A. 

The distribution of helicity injection by flux emergence and shuffling motions in Figure 2h and 2i demonstrate the origin of positive helicity injection. From both panels, we see that the positive helicity injection along the neutral line comes from $\dot{H_{n}}$, while, the other positive injection comes from $\dot{H_{t}}$ at the region A. The complex evolution properties along the neutral line was studied by several authors (Wang et al., 2008; Park et al., 2010), and was conjectured as a result of the emergence of a positive helicity system. In region A, the positive injection of the helicity in $\dot{H_{t}}$ map is consistent with the clockwise rotation of sunspot. While in $\dot{H_{n}}$ map, the negative helicity injection is consistent with the left handed twist of magnetic field.         
 
The temporal evolution of $\dot{H}$ on December 11 was shown in Figure 3. Triangles indicate the contribution of flux emergence, stars represent the shuffling motions, and diamonds indicate the sum of both. The top panel shows that the flux emergence and shuffling motions contribute almost the same sign of magnetic helicity during whole period. The helicity injected by the flux emergence is almost negative which is consistent with the dominated left handed twist of the magnetic field for south-hemisphere AR. On the other hand, helicity injected by shuffling motion varies from time to time. It is negative until 12:00 UT and then becomes positive. The dynamical transient variation of $\dot{H}$ is consistent with $\dot{H}_{t}$, while the variation of $\dot{H}_{n}$ is a little bit stable. The variation tendency derived by us is almost the same as that by Park et al. (2010), although the exact value was not equal because of the differences of data type, observational time and calculation method.

Middle and bottom panels show the temporal evolution of $\dot{H_{t}}$ and $\dot{H_{n}}$ for following and leading sunspot, respectively. For the following sunspot, the helicity injections by the shuffling motion and flux emergence are always negative, and the helicity accumulation by both effects are $-178\times10^{40} Mx^{2}$ and $-164\times10^{40} Mx^{2}$, respectively. For the leading sunspot, the helicity injections caused by both effects are negative at first, and then associated with the flux emergence along the neutral line and the clockwise motion in the region A, it changes its sign from negative to positive. The helicity accumulation by the shuffling motion and flux emergence are $248\times10^{40} Mx^{2}$ and $46\times10^{40} Mx^{2}$, respectively. Here it is worth noting that the magnetic helicity accumulation in this active region is negative. And the period which we studied includes a positive helicity transport phase (Park et al., 2010). Even though as a whole, the magnetic helicity accumulation in this active region is $-48\times10^{40} Mx^{2}$.  Our results show that the helicity accumulation mainly result from the following sunspot, but the dynamic variation mainly result from the leading sunspot. And also, even though the magnetic flux of the following sunspot is much smaller than that of the leading sunspot, it contributes most helicity accumulation because of the predominance of the negative helicity injection to the solar corona in this AR. Moreover the active region was located in the southern hemisphere and showed the left-handed (negative) helicity, violating the so-called hemispheric rule. So the characteristics of helicity evolution of the region may be specific to that category of ARs, which should be studied by further observations.

\section{Summary and Discussion}

By fully using the observational data, we have developed a new methodology by which we can estimate the magnetic helicity injection by flux emergence and shuffling motions, separately. The key point of the methodology is that we deduce the component of plasma velocity, which is perpendicular to the magnetic field, by observational data only. It provide us a useful tool to study the contribution of flux emergence and shuffling motion in energy storage and eruption initiation, especially after the launch of SDO. The derivation of the helicity injection is demonstrated for AR NOAA 10930. The observational properties of this AR during the period of our study are: a stable mature leading sunspot and a fast emerging and rapidly rotating following sunspot. Some important conclusions are obtained:   

1. The sign of the helicity fluxes from flux emergence and shuffling motions are the same. The temporal variations of $\dot{H}_n$ is relatively stable, while $\dot{H}_t$ changes rapidly. The variation tendency of $\dot{H}$ is consistent with $\dot{H}_t$, implying that flux emergence provided the continuously accumulation of the helicity which may play a key role in helicity or energy storage, while the shuffling motions added some dynamic change of the helicity injection which may play a key role in eruption initiation. 

2. For the following sunspot, helicity injection from flux emergence and shuffling motion are all negative. Helicity flux from flux emergence is relatively stable and compared with that from shuffling motions.

3. For the leading sunspot, helicity injection changes its sign from negative to positive for both effects, which is conjectured associated with emergence of a magnetic flux which contains opposite helicity and a clockwise rotation in southwest of the sunspot.

Usually, the rotational motion and the twisted magnetic field in sunspot occur simultaneously. We always want to know whether the observed rotation of sunspots represents the emergence of twisted magnetic flux tube, or the rotational shuffling motion twists of the magnetic flux tube. When the rotational motion represents the emergence of a twisted flux rope, one can expect that the motion results from the emergence will along the magnetic field. However in AR 10930, we find the rotational motion remains in $\mbox{\boldmath$\upsilon$}_{\perp t}$ map. So we conjecture that the rotational motion in AR NOAA 10930 is the real shuffling motion, although more observational evidence will be necessary for the final conclusion to the question stated above. 

\acknowledgments
Y. Zhang wishes to address her sincere thanks to Dr. C. L. Jin for providing her code to revise the SP doppler velocity. We are grateful to the Hinode team for providing the wonderful data. Hinode is a Japanese mission developed and launched by ISAS/JAXA, with NAOJ as domestic partner and NASA and STFC (UK) as international partners. It is operated by these agencies in co-operation with ESA and NSC (Norway).

\clearpage


\clearpage

\end{document}